\documentclass[aps,prb,reprint,groupedaddress,showpacs,amsfonts,amsmath,amssymb,superscriptaddress]{revtex4-1}
\usepackage{graphicx}
\usepackage{bm}
\usepackage{color}
\begin{document}
\title{Dynamics of a single hole in the Heisenberg-Kitaev model: a self-consistent Born approximation study}
\author{Bin-Bin Wang}
\affiliation{National Laboratory of Solid State Microstructures and Department of Physics, Nanjing University, Nanjing 210093, China}
\author{Wei Wang}
\affiliation{National Laboratory of Solid State Microstructures and Department of Physics, Nanjing University, Nanjing 210093, China}
\author{Shun-Li Yu}
\email{slyu@nju.edu.cn}
\affiliation{National Laboratory of Solid State Microstructures and Department of Physics, Nanjing University, Nanjing 210093, China}
\affiliation{Collaborative Innovation Center of Advanced Microstructures, Nanjing University, Nanjing 210093, China}
\author{Jian-Xin Li}
\email{jxli@nju.edu.cn}
\affiliation{National Laboratory of Solid State Microstructures and Department of Physics, Nanjing University, Nanjing 210093, China}
\affiliation{Collaborative Innovation Center of Advanced Microstructures, Nanjing University, Nanjing 210093, China}
\date{\today}
\begin{abstract}
The magnetic properties of $4d$ and $5d$ transition-metal insulating compounds with the honeycomb structure are believed to be described by the Heisenberg-Kitaev model, which contains both the isotropic Heisenberg interaction $J$ and anisotropic Kitaev interaction $K$. In this paper, to investigate the charge dynamics in these materials, we study the single-hole propagation of the $t$-$J$-$K$ model in various magnetically ordered phases by the self-consistent Born approximation. We find that there are low-energy coherent quasiparticle (QP) excitations in all of these phases which appear firstly around the $K$ point in the Brillouin zone (BZ), but the band-widths of these QPs are very small due to the hole-magnon coupling.  Interestingly, in the zigzag phase relevant to recent experiments, though the QP weights are largely suppressed in the physical spectra in the first BZ, we find that they recover in the extended BZs. Moreover, our results reveal that the low-energy QP spectra are reduced with the increase of $K$.
\end{abstract}

\maketitle

\section{INTRODUCTION}

Since the discovery of high-Tc superconducting cuprates, the nature of the charge carrier in doped Mott insulators has attracted considerable attention in the studies of strongly correlated electron systems\cite{RevModPhys.78.17}. The parent compounds of cuprates are antiferromagnets (AF) Mott insulators whose physics could be described by a Hubbard model with large on-site Coulomb repulsion $U$. At half filling, the Hubbard model reduces to the AF Heisenberg model\cite{science.235.4793.1196}. By doping holes or electrons into the system, the AF order is suppressed and a superconducting phase emerges above a critical doping concentration, and the low-energy physics in this case is believed to be described by the $t$-$J$ model\cite{PhysRevB.37.3759}. The evolution from the AF insulating phase to the superconducting phase induced by doping is highly nontrivial\cite{RevModPhys.78.17}. The study of charge dynamics in Mott insulators is essential to understand the extraordinary phenomena in cuprates. In this respect, the dynamics of a single hole or electron in an AF Mott insulator on the square lattice is an outstanding issue, and it has been extensively studied\cite{PhysRevLett.60.2793,PhysRevB.39.6880,PhysRevB.43.10882,PhysRevB.41.9049,PhysRevB.44.2414,PhysRevLett.73.728,PhysRevB.55.5983,PhysRevB.90.245102}.

Recently, the $4d$ and $5d$ transition-metal materials have attracted considerable attentions, as the interplay between the spin-orbital coupling (SOC), crystal fields, and electronic correlation can induce many novel electronic and magnetic ground states\cite{annurev-conmatphys.5.57,annurev-conmatphys.7.195}. Especially, it could lead to the so-called spin-orbital assisted Mott insulator, in which the relevant electronic structures are described in terms of a half-filled relativistic $J_{eff}=1/2$ narrow band so that a small Hubbard interaction $U$ is sufficient to open a Mott gap\cite{PRL.101.076402,science.323.1329}. The low-energy physics of such insulators is described in terms of $J_{eff}=1/2$ pseudospin Hamiltonians. In particular, the Kitaev interactions that underlying the celebrated Kitaev honeycomb model\cite{Ann.Phys.321.2} can be realized in such insulators on the honeycomb lattice\cite{PRL.102.017205,PRL.105.027204}, such as Na$_{2}$IrO$_{3}$, Li$_{2}$IrO$_{3}$ and $\alpha$-RuCl$_{3}$\cite{PRL.108.127204,PRL.110.076402,PRB.90.041112,PRL.114.147201,Nat.Mater.15.733,PRL.118.107203,PRB.93.155143,PRB.93.214431,PRB.96.115103}. The Kitaev model is exactly solvable and its ground state is a $Z_{2}$ spin liquid whose elementary excitations are Majorana fermions\cite{Ann.Phys.321.2}. However, in real materials, besides the Kitaev interactions, there are also some other types of interactions\cite{PRL.102.017205,PRL.105.027204,PRL.108.127204,PRL.110.076402,PRB.90.041112,PRL.114.147201,Nat.Mater.15.733,PRL.118.107203,PRB.93.155143,PRB.93.214431,PRB.96.115103}, such as the Heisenberg interactions. As a result, in most cases, the real materials are magnetically ordered, e.g. the ground states of Na$_{2}$IrO$_{3}$ and $\alpha$-RuCl$_{3}$ exhibit the zigzag magnetic order. Based on the Heisenberg-Kitaev model containing both the nearest-neighbor (NN) Heisenberg and Kitaev exchange interactions, theoretical studies have shown that the zigzag AF phase emerges in a broad range of parameters\cite{PhysRevLett.110.097204}. Moreover, besides the zigzag AF phase, there are other three magnetically ordered phases in this model in the global phase diagram\cite{PhysRevLett.110.097204}, including a ferromagnetic (FM) order, a N\'{e}el AF order and a stripy order. Experimentally, the angle-resolved photoemission spectroscopy (ARPES) measurements on Na$_{2}$IrO$_{3}$ and $\alpha$-RuCl$_{3}$\cite{PhysRevLett.109.266406,PhysRevB.94.161106,PhysRevLett.117.126403,srep39544,PhysRevMaterials.1.052001} show that the small bandwidths of the Ir $5d$-$t_{2g}$ and Ru $4d$-$t_{2g}$ valence bands are inconsistent with the large hopping amplitudes for $5d$-$t_{2g}$ and $4d$-$t_{2g}$ states as generally expected. Hence,
the interplay between magnetism and charge dynamics is important in the theoretical understanding of the spectral properties in these materials\cite{PhysRevLett.109.266406,PhysRevB.94.161106}.

In this paper, motivated by the progress in research of the $4d$ and $5d$ transition-metal compounds and the possible applications of the Heisenberg-Kitaev model in these materials, we investigate the dynamics of a single hole in various magnetically ordered phases of this model. We find  there are low-energy coherent quasiparticle (QP) excitations with small bandwidths in all of these phases, though the spectra at high energy are dominated by large incoherent spectral weights. The small bandwidths of the QP bands are resulted from the strong hole-magnon couplings. We also find that the low-energy coherent QPs in all of these phases appear firstly around the $K$ point in the BZ, which suggests that the doped hole will form a hole Fermi pocket centered at this point for small doping levels. Interestingly, for the zigzag phase that is relevant to Na$_{2}$IrO$_{3}$ and $\alpha$-RuCl$_{3}$, clear QP features appear in spectral functions of holes created and annihilated on one sublattice, while most of them are hidden in the physical spectral functions in the first BZ due to the interference effect of the two-sublattice Green's function on the honeycomb lattice. This interference effect also manifests itself in the way of recovering these hidden spectral weights in the extended BZs. Moreover, when the Kitaev interaction is increased to drive the system close to the Kitaev spin-liquid phase, the low-energy QP spectral weights are largely suppressed. The physical mechanisms of inducing these spectral features are also discussed in this paper.

\section{MODEL AND METHOD}

\begin{figure}
  \centering
  \includegraphics[scale=0.54]{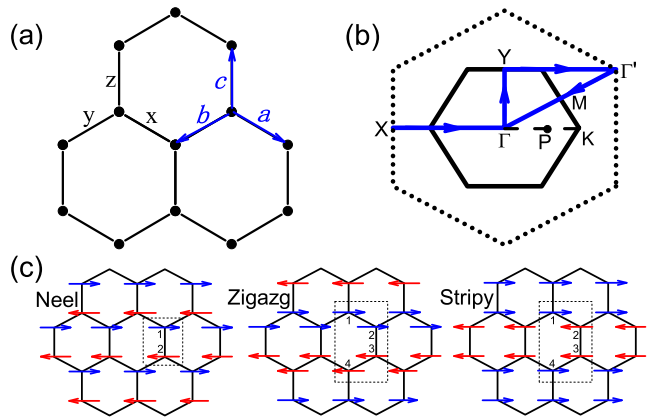}
  \caption{\label{fig:hrbz}(color online) (a) Three different directions (labeled by $x$, $y$ and $z$) of the NN bonds on the honeycomb lattice, and the corresponding vectors are represented by $\bm{a}$, $\bm{b}$ and $\bm{c}$. (b) First BZ (black solid lines) with high symmetric points. The blue lines with arrow indicate the path along the symmetric directions, and the black dotted lines connect the $\Gamma^{\prime}$ points in the extended BZs. (c) Magnetic structures of the N\'{e}el AF, zigzag and stripy phases. The dashed rectangles show the magnetic unit cells.}
\end{figure}

Our analysis is based on the $t$-$J$-$K$ model which consists of two terms,
\begin{equation}
H=H_{t}+H_{JK},
\label{eq:h}
\end{equation}
where $H_{JK}$ is the Hamiltonian of the Heisenberg-Kitaev model and the hopping term $H_{t}$ is restricted in the Hilbert space without double occupancies. The two terms are given as
\begin{equation}
H_{t}=t\sum_{\langle ij\rangle\sigma}c^{\dag}_{i\sigma}c_{j\sigma},
\label{eq:ht}
\end{equation}
\begin{equation}
H_{JK}=\sum_{\langle ij\rangle}(J\bm{S}_{i}\cdot\bm{S}_{j}+KS_{i}^{u_{ij}}S_{j}^{u_{ij}}),
\label{eq:h0}
\end{equation}
where $c^{\dag}_{i\sigma}$ is the electron creation operator with spin $\sigma$, $\bm{S}_{i}$ is the electron spin operator and the index $u_{ij}$ takes values $x$, $y$, or $z$ depending on the direction of the NN bond $\langle ij\rangle$ [see figure~\ref{fig:hrbz}(a)].

We will employ the self-consistent Born approximation (SCBA) to investigate the charge dynamics in magnetically ordered phases of the $t$-$J$-$K$ model. In SCBA, the electron operators in equation~(\ref{eq:ht}) are expressed by the slave-fermion representation\cite{PhysRevB.39.6880}, $c_{i\sigma}=h^{\dag}_{i}b_{i\sigma}$, where $h^{\dag}_{i}$ is a fermionic operator creating a spinless hole (holon) and $b_{i\sigma}$ is a Schwinger boson operator annihilating a boson with spin $\sigma$ at the site $i$. The fermionic and bosonic operators satisfy the following constraint:
\begin{equation}
h^{\dag}_{i}h_{i}+\sum_{\sigma}b^{\dag}_{i\sigma}b_{i\sigma}=1.
\end{equation}
The spin operators in equation~(\ref{eq:h0}) are expressed as $\bm{S}_{i}=\chi_{i}^{\dag}\bm{\sigma}\chi_{i}$, where $\chi^{\dag}_{i}=(b^{\dag}_{i\uparrow},b^{\dag}_{i\downarrow})$ is a two-component spinor and $\bm{\sigma}$ represents the Pauli matrices.
For an ordered state, one of the bosons condenses and the remaining bosonic operators are described by the Holstein-Primakoff bosonic operators, subsequently the spin excitations can be calculated by the linear spin-wave (LSW) theory\cite{PRB.96.115103}. Then, we determine how a hole couples to these spin excitations when it hops. This approach has been widely applied to study a single hole dressed by spin excitations in various magnets\cite{PhysRevLett.60.2793,PhysRevB.39.6880,PhysRevB.43.10882,PhysRevB.53.402,PhysRevB.54.13158,PhysRevB.73.155118,PhysRevB.84.195125}.

Assuming there are $L$ sublattices in each magnetic unit cell, we can obtain the following effective spin-wave Hamiltonian from the Hamiltonian (\ref{eq:h0}) with the LSW theory\cite{PRB.96.115103},
\begin{equation}
H_{LSW}=\frac{1}{2}\sum_{k}\bm{X}^\dag_{k}\bm{M}(k)\bm{X}_{k},
\label{h_lsw}
\end{equation}
where $\mathbf{X}^{\dag}_{k}=(b^{\dag}_{k,1},\cdots,b^{\dag}_{k,L},b_{-k,1},\cdots,b_{-k,L})$ with $b_{k,\alpha}=\sqrt{\frac{L}{N}}
\sum_i{e^{-i\bm{k}\cdot\bm{r}_{i\alpha}}b_{i\alpha}}$. Here, $b_{i\alpha}$ is the Holstein-Primakoff boson on the sublattice $\alpha$ of the $i$-th magnetic unit cell, and $N$ is the total number of lattice sites. The explicit forms of $2L\times2L$ matrix $\bm{M}(k)$ depend on the magnetic orders of ground states, and are given in~\ref{sw-h}. By diagonalizing the Hamiltonian (\ref{h_lsw}), we have
\begin{equation}
H_{LSW}=\sum_{kn}{\omega_{k,n}\gamma_{k,n}^\dag  \gamma_{k,n}},
\label{eq:hdiagn}
\end{equation}
where $\gamma_{k,n}^\dag$ and $\gamma_{k,n}$ are the creation and annihilation operators for the $n$-th magnon mode with energy $\omega_{k,n}$.
The spin-wave dispersions for several typical interaction parameters used in this paper are shown in~\ref{spin-wave}.

After replacing the Schwinger bosons by the Holstein-Primakoff bosons, the electron operators now can be rewritten as: ${c}_{i\alpha\downarrow}=h_{i\alpha}^{\dag}b_{i\alpha}$ and ${c}_{i\alpha\uparrow}=h_{i\alpha}^\dag$ for the local moment along the $+z$ direction, and ${c}_{i\alpha\uparrow}=h_{i\alpha}^{\dag}b_{i\alpha}$ and ${c}_{i\alpha\downarrow}=h_{i\alpha}^\dag$ for the local moment along the $-z$ direction.
Using the Fourier transformation
$h_{k,\alpha}=\sqrt{\frac{L}{N}}\sum_{\bm{k}}e^{-i\bm{k}\cdot\bm{r}_{i\alpha}}h_{i\alpha}$,
the Hamiltonian $H_t$ can be written as
\begin{eqnarray}\label{hole-eq}
H_{t}&=\sum_{k}\bm{\Psi}_{k}\bm{T}(k)\bm{\Psi}^{\dag}_{k}
+\sum_{kqn}\bm{\Psi}_{k}\left[\gamma^{\dag}_{q,n}\bm{D}(k,q,n)\right. \nonumber \\
&\left.+ \gamma_{-q,n}\bm{D}^{\dag}(k-q,-q,n)\right]\bm{\Psi}^{\dag}_{k-q},
\end{eqnarray}
where $\bm{\Psi}_{k}=(h_{k,1},h_{k,2},\cdots,h_{k,L})$. The bare holon hopping matrix $\bm{T}(k)$ and holon-magnon interaction vertex $\bm{D}(k,q,n)$ are $L\times L$ matrices, and their expressions depend on the magnetic orders (see~\ref{hole-magnon-vertex}).

The holon Green's function is written as,
\begin{equation}
\bm{G}^{h}(k,\omega)=\left[\omega-\bm{T}(k)-\bm{\Sigma}^{h}(k,\omega)\right]^{-1}.
\label{eq:gh}
\end{equation}
In the SCBA, the renormalized vertex and magnon propagator in the self-energy are approximated by the bare vertex and propagator, so the self-energy matrix $\bm{\Sigma}^{h}(k,\omega)$ is given as,
\begin{equation}
\bm{\Sigma}^{h}(k,\omega)=\sum_{qn}\bm{D}(k,q,n)
\bm{G}^{h}(k-q,\omega-\omega_{q,n})\bm{D}^{\dag}(k,q,n).
\label{eq:sigma}
\end{equation}
The corresponding holon spectral function is
\begin{equation}
A^{h}_{\alpha\beta}(k,\omega)=-\frac{1}{\pi}\rm{Im} G^{h}_{\alpha\beta}(k,\omega).
\end{equation}

As the holes from different sublattices can not be distinguished in experiments, in order to compare with ARPES experiments we have to introduce the operator $c_{k,\sigma}$ by the Fourier transformation\cite{PhysRevB.56.11769}
$c_{k,\sigma}=\sqrt{\frac{2}{N}}\sum_{i\alpha}{c_{i\alpha\sigma}e^{i\bm{k}\cdot\bm{r}_{i\alpha}}}$. The corresponding Green's function $G_{\sigma\sigma^{\prime}}^{c}(k,\omega)$ is defined as
\begin{equation}
G_{\sigma\sigma^{\prime}}^{c}(k,\omega)=-i\int_0^\infty\langle0|c_{k,\sigma}^{\dag}(t)c_{k,\sigma^{\prime}} (0)|0\rangle e^{i\omega t}dt,
\end{equation}
and the spectral function for the physical hole is
\begin{equation}
A^{c}_{\sigma\sigma^{\prime}}(k,\omega)=-\frac{1}{\pi}\rm{Im}G_{\sigma\sigma^{\prime}}^{c}(k,\omega).
\end{equation}
$\bm{G}^{c}(k,\omega)$ can be calculated from $\bm{G}^{h}(k,\omega)$ with the following relation\cite{PhysRevB.56.11769},
\begin{widetext}
\begin{align}
\bm{G}^{c}(k,\omega)&=\bm{A}^{\dag}(k)\bm{G}^{h}(k,\omega)\bm{A}(k)
+\sum_{qn}\bm{B}^{\dag}(k,q,n)\bm{G}^{h}(k-q,\omega-\omega_{q,n})\bm{B}(k,q,n) \nonumber \\
&+\sum_{qn}\bm{A}^{\dag}(k)\bm{G}^{h}(k,\omega)\bm{D}(k,q,n)\bm{G}^{h}(k-q,\omega-\omega_{q,n})\bm{B}(k,q,n) \nonumber \\
&+\sum_{qn}\bm{B}^{\dag}(k,q,n)\bm{G}^{h}(k-q,\omega-\omega_{q,n})\bm{D}^{\ast}(k,q,n)\bm{G}^{h}(k,\omega)\bm{A}(k) \nonumber \\
&+\sum_{qq^{\prime}nn^{\prime}}\bm{B}^{\dag}(k,q,n)\bm{G}^{h}(k-q,\omega-\omega_{q,n})\bm{D}^{\ast}(k,q,n)
\bm{G}^{h}(k,\omega)\bm{D}(k,q^{\prime},n^{\prime})\bm{G}^{h}(k-q^{\prime},\omega-\omega_{q^{\prime},n^{\prime}})
\bm{B}(k,q^{\prime},n^{\prime}),
\end{align}
\end{widetext}
where $A_{\alpha\sigma}(k)=\langle0|h_{k,\alpha}c_{k,\sigma}|0\rangle$ and $B_{\alpha\sigma}(k,q,n)=\langle0|h_{k-q,\alpha}\gamma_{n,q}c_{k,\sigma}|0\rangle$.

\section{results and discussion}

In the following, we analyze the spectral properties of a single hole in various magnetically ordered phases of this model, including the N\'{e}el AF, zigzag and stripy phases, whose magnetic structures are shown in figure~\ref{fig:hrbz}(c). In the calculations, we choose $16\times16$ magnetic unit cells and use the periodic boundary condition. The $\omega$ mesh is set to $2000$ points from $-6t$ to $6t$. Roughly speaking, for Mott insulators, $J\propto t^{2}/U$, so we set $J$ much smaller than $t$ in the following discussions.

\begin{figure}
  \centering
  \includegraphics[scale=1.1]{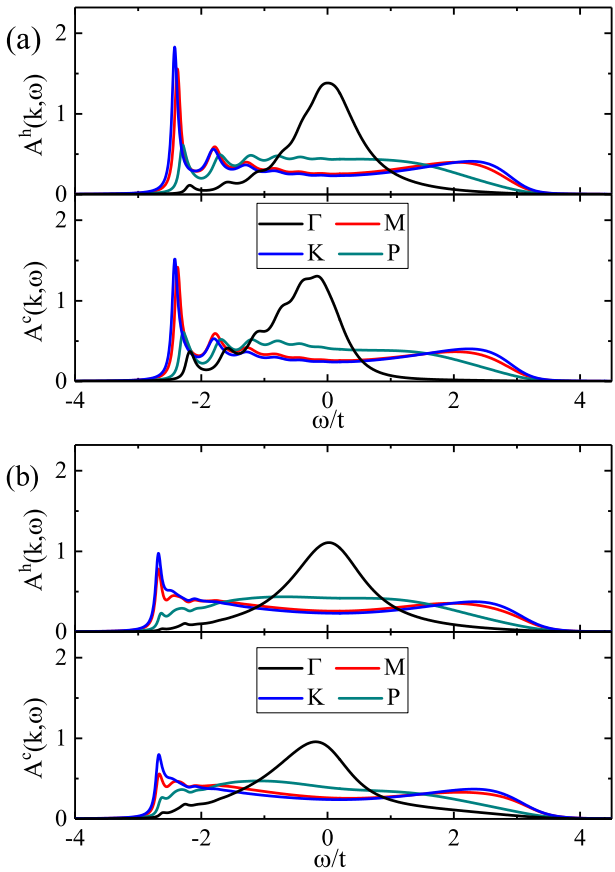}
  \caption{\label{fig:neel}(color online) Spectral functions $A^{h}(\bm{k},\omega)$ and $A^{c}(\bm{k},\omega)$ of a single hole at several high symmetric points $(\Gamma,{\rm M,K,P})$ [see figure~\ref{fig:hrbz}(b)] of the BZ in the N\'{e}el AF phase for (a) $J=0.16t$ and $K=-0.08t$, (b) $J=0.035t$ and $K=0.40t$.}
\end{figure}
We first discuss the case of the N\'{e}el AF phase. According to the results in reference~\cite{PhysRevLett.110.097204}, the N\'{e}el AF phase exists in the range $-0.7355\lesssim J/K\lesssim0.022$ with $J>0$. Figure~\ref{fig:neel}(a) shows the holon spectra $A^{h}(k,\omega)=\sum_{\alpha}A^{h}_{\alpha\alpha}(k,\omega)$ and the spectra for the physical hole $A^{c}(k,\omega)=\sum_{\sigma}A^{c}_{\sigma\sigma}(k,\omega)$ as defined above for $J=0.16t$ and $K=-0.08t$. In addition to the dominant incoherent spectral weight at high energies, we can find that there are obvious quasiparticle (QP) coherent peaks with a small bandwidth near the bottom of the whole spectra. In the N\'{e}el AF phase, a hole moving along a chain (zigzag or armchair) will destroy the antiferromagnetic alignment of spins and create a string of flipped spins with an increase in energy being proportional to the length of the path. Subsequently, the hole tends to be bounded to its original lattice site by the string, so it seems to be immobile. This is consistent with our approximation, in which the hopping term of the free holon is zero (see~\ref{hole-magnon-vertex}). However, Trugman has suggested that there are certain higher-order hopping processes\cite{PhysRevB.37.1597}, which allow the hole to move onto a next-nearest-neighbor site without creating frustration. On the other hand, the quantum spin fluctuations due to the $S^{+}S^{-}$ and $S^{+}S^{+}$ terms in the Kitaev-Heisenberg model (\ref{eq:h0}) can erase part of the string and make the hole mobile. Thus, the holon (spinless hole) does not have an infinite effective mass but has a finite mobility, which exhibits an obvious dispersion for the QP peak as shown in figure~\ref{fig:neel}(a). These spectra also reveal that the lowest-energy QP appears at the $K$ point and disperses to the small momentum $P$ point and nearly loses its weight at the $\Gamma$ point.

From figure~\ref{fig:neel}(a), we can see that $A^{h}(k,\omega)$ and $A^{c}(k,\omega)$ have little difference except that the QP peak is slightly suppressed in the physical spectral function $A^{c}(k,\omega)$. Furthermore, in comparison with the exact-diagonalization (ED) results\cite{PRB.90.024404}, we find that both the overall shapes of the spectra and the dispersion of the lowest qusiparticles are very similiar.

To see the effects of the Kitaev interaction $K$ on the spectra, we change the interaction parameters to $J=0.035t$ and $K=0.40t$, and the results are shown in figure~\ref{fig:neel}(b). In comparison with the results in figure~\ref{fig:neel}(a), we find that the line shapes of the spectra look similar in the two sets of parameters. The main difference is that the low-energy QP spectral weight is suppressed significantly in figure~\ref{fig:neel}(b) and it transfers to the high-energy incoherent part. In particular, the QP peaks are smeared out at the $\Gamma$ and $P$ points. The reason for this reduction of QP spectral weight is that the energy of the low-energy branch of the magnons decreases with the raise of $K$ [see figure~\ref{fig:spin-wave}(a)], which enhances the coupling between the holon and magnon and correspondingly reduces the coherence of the QPs.

\begin{figure}
  \centering
  \includegraphics[scale=1.1]{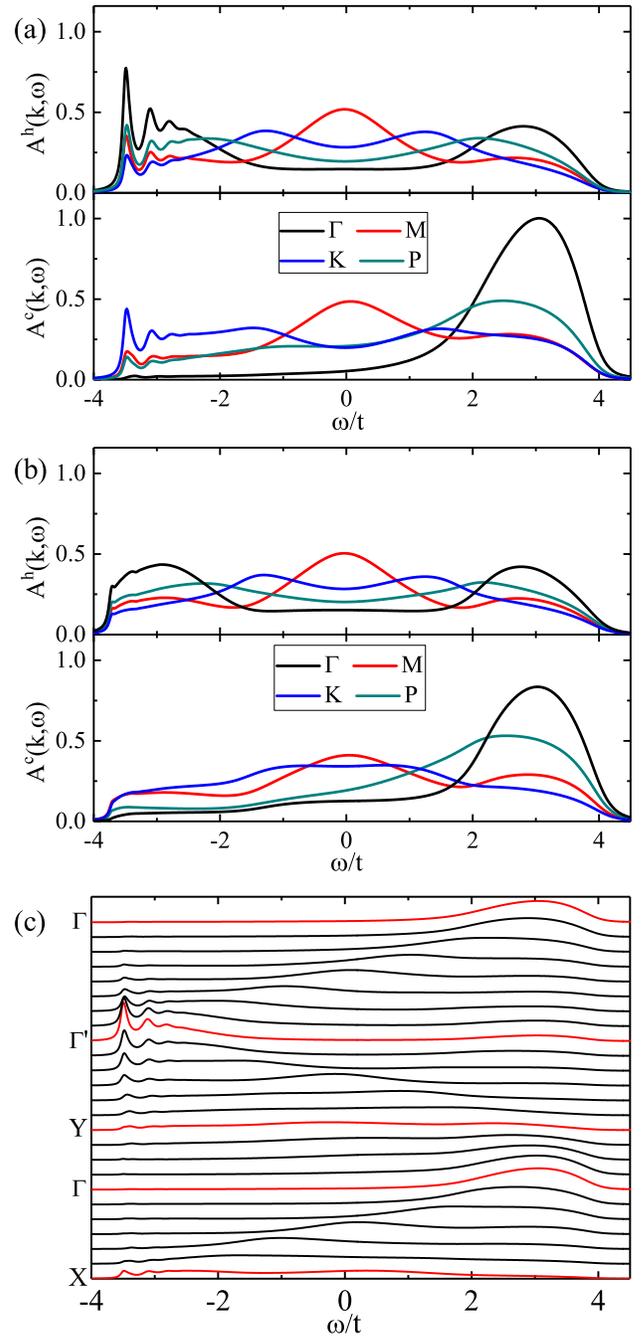}
  \caption{\label{fig:zigzag}(color online) Spectral functions of a single hole in the zigzag phase. $A^{h}(\bm{k},\omega)$ and $A^{c}(\bm{k},\omega)$ at several high symmetric points of the BZ for (a) $J=-0.09t$ and $K=0.22t$, (b) $J=-0.017t$ and $K=0.4t$. (c) $A^{c}(\bm{k},\omega)$ along the high symmetric lines of the BZ [see figure~\ref{fig:hrbz}(b)].}
\end{figure}
We then study the single-hole spectra in the zigzag phase. The parameter range for the zigzag phase is $-1.4909\lesssim J/K\lesssim-0.0252$ with $J<0$ (see reference~\cite{PhysRevLett.110.097204}). Figure~\ref{fig:zigzag}(a) shows $A^{h}(k,\omega)$ and $A^{c}(k,\omega)$ for $J=-0.09t$ and $K=0.22t$. As the original point-group symmetry of the honeycomb lattice is broken in the zigzag phase, we average the spectral functions over all of the inequivalent $K$ ($M$ or $P$) points. In contrast to the N\'{e}el AF phase, the hopping Hamiltonian of the free holon in the zigzag phase is nonzero (see \ref{hole-magnon-vertex}), i.e. the holon can hop along the zigzag chains. However, due to the strong coupling between the holon and the spin waves of localized spins in the magnetic background, the well-defined QP only exists in the low-energy region. Thus, as shown in figure~\ref{fig:zigzag}(a), the bandwidth of the QP is largely suppressed, which makes the QP spectra exhibit very weak dispersion. This mechanism can be further verified by increasing $K$ (or reducing $J$), for which the magnon energy of the low-energy branch is decreased [see figure \ref{fig:spin-wave}(b)] and correspondingly the coupling between the holons and magnons is enhanced. The enhancement of the holon-magnon coupling will completely suppress the QP spectral weight and there is no QP peak in all $k$ points[see figure~\ref{fig:zigzag}(b)]. This spin-polaronic behavior in the spectral function provides a natural explanation for the ARPES measurements on Na$_{2}$IrO$_{3}$ and $\alpha$-RuCl$_{3}$\cite{PhysRevLett.109.266406,PhysRevB.94.161106,PhysRevLett.117.126403,srep39544,PhysRevMaterials.1.052001}, which show that the small bandwidths of the Ir $5d$-$t_{2g}$ and Ru $4d$-$t_{2g}$ valence bands are at variance with the generally expected large hopping amplitudes for $5d$-$t_{2g}$ and $4d$-$t_{2g}$ states.

Unlike the N\'{e}el AF phase, the spectra $A^{h}(k,\omega)$ and $A^{c}(k,\omega)$ have distinct features in the zigzag phase. For $A^{h}(k,\omega)$, we see QP peaks very clearly at low energies. However, for $A^{c}(k,\omega)$, the QP features are obviously suppressed at the $\Gamma$ and $P$ points, and there is even no QP peak at the $\Gamma$ point, while the spectral intensity at the $K$ point is slightly enhanced. Thus, for the physical spectral function the QP peak also occurs firstly at the $K$ point. These features are similar to the ED results\cite{PhysRevLett.111.037205}. The hopping processes of the hole in the zigzag phase mainly come from the NN bonds between the two sublattices of the honeycomb lattice. As a result, if the intra-sublattice and inter-sublattice spectral functions have comparable intensity but different signs at some momentum points, the corresponding spectra are seriously suppressed, otherwise the spectra will be enhanced if they have the same sign.

On the other hand, the two-sublattice structure introduces a phase difference upon a translation of the reciprocal primitive vector\cite{EPL.115.27008}, which results in a larger periodic unit cell of the spectral function $A^{c}(k,\omega)$. Correspondingly, the low-energy spectral weights of $A^{h}(k,\omega)$ hidden in $A^{c}(k,\omega)$ will recover in extended BZs [see figure~\ref{fig:zigzag}(c)], so we propose that the ARPES experiments will find more information about the spectral function in extended BZs.

\begin{figure}
  \centering
  \includegraphics[scale=1.1]{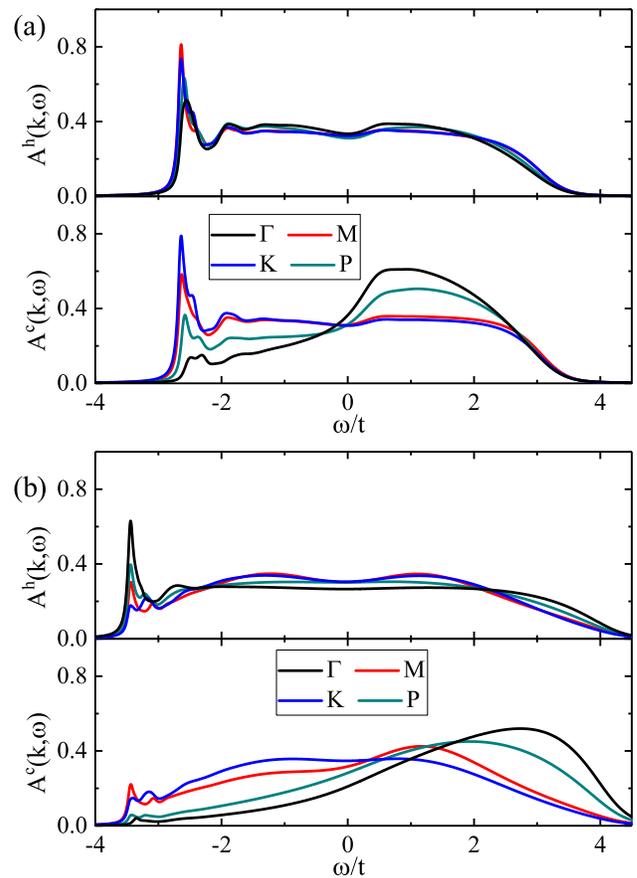}
  \caption{\label{fig:stripy}(color online) Spectral functions $A^{h}(\bm{k},\omega)$ and $A^{c}(\bm{k},\omega)$ of a single hole at several high symmetric points of the BZ in the stripy phase for (a) $J=0.16t$ and $K=-0.24t$, (b) $J=0.068t$ and $K=-0.38t$.}
\end{figure}
Next, we shift to the stripy phase, which exists in the range of $-0.7332\lesssim J/K\lesssim-0.1253$ with $J>0$ (see reference~\cite{PhysRevLett.110.097204}). Figure~\ref{fig:stripy}(a) shows $A^{h}(k,\omega)$ and $A^{c}(k,\omega)$ for $J=0.16t$ and $K=-0.24t$. Similar to the zigzag phase, we also average the spectral functions over the inequivalent points $K$, $M$ and $P$. A notable feature of $A^{h}(k,\omega)$ is that the spectra at different momenta are very similar, so the QP dispersion is absent in the stripy phase. The reason is that the alternating alignment of the AF and FM bonds hinders the coherent motion of the hole, which is similar to the case of the N\'{e}el AF phase. Moreover, the Trugman processes\cite{PhysRevB.37.1597}, which allow the hole to move onto a third-nearest-neighbor site without creating frustration in the stripy phase, need higher-order corrections compared to the N\'{e}el AF phase. Thus, the Trugman processes can not induce an obvious dispersion of the QPs as that in the N\'{e}el AF phase. In addition, similar to the zigzag phase, the low-energy spectra at $\Gamma$ and $P$ points in $A^{c}(k,\omega)$ are seriously suppressed comparing with those in $A^{h}(k,\omega)$, and the physical mechanism is the same as that in the zigzag phase. Also, similar to those in the N\'{e}el AF and zigzag phases, the increase of the Kitaev interaction $K$ enhances the hole-magnon coupling and suppresses the low-energy QP spectral weight [see figure \ref{fig:stripy}(b)].

At last, comparing the above results about the physical spectral function $A^{c}(k,\omega)$, we find a common feature of the low-energy QPs, i.e. the lowest excitation energy of the QPs is at the $K$ point. It suggests that the doped hole will appear firstly around the $K$ point, and consequently form a hole Fermi pocket centered at this point in the small doping regime. This feature is consistent with the ARPES results on Na$_{2}$IrO$_{3}$ and $\alpha$-RuCl$_{3}$\cite{PhysRevLett.109.266406,PhysRevB.94.161106,PhysRevLett.117.126403}. Moreover, in all of the three magnetically ordered phases, the low-energy QP spectral weights are suppressed by the increase of the Kitaev interaction $K$, for which the phases are close to the Kitaev spin-liquid phase. This result is consistent with the ED result, which shows that the QP spectral weight disappears in the Kitaev spin-liquid phase\cite{PRB.90.024404}.

\section{Summary}

We have studied charge dynamics of a single hole in various magnetically ordered phases of the Kitaev-Heisenberg model by the self-consistent Born approximation. Though the spectra are dominated by large incoherent spectral weight, there are low-energy coherent QP excitations in all of these phases. We find that the doped hole appears firstly around the $\Gamma$ point in the Brillouin zone, suggesting the formation of a hole Fermi pocket around that momentum point with a light doping.
The spectra are modified remarkably when increasing Kitaev interaction drives the system close to the Kitaev spin-liquid phase, i.e. the QP features are strongly suppressed and the spectral weight moves to high energy. Interestingly, in the zigzag phase, clear QP features appear in spectral functions in the first BZ for holes created and annihilated on one sublattice, while they are hidden in the physical spectral functions corresponding to the ARPES experiments, but we find that these hidden spectral recovers in the extended BZs. These results may stimulate further experimental investigations on dynamics of a single hole, especially in candidate Kitaev-Heisenberg materials with an antiferromagnetic and stripy magnetic order.

\begin{acknowledgments}
This work was supported by the National Natural Science Foundation of China (Grants No. 11674158,
and No. 11774152), National Key Projects for Research and Development of China (Grant No. 2016YFA0300401), and Natural Science Foundation of Jiangsu Province (BK20140589). W.W. was also
supported by the program B for Outstanding PhD candidate of Nanjing University.
\end{acknowledgments}

\appendix
\renewcommand{\thefigure}{\thesection\arabic{figure}}

\section{Explicit expressions of spin-wave Hamiltonian}\label{sw-h}

The hopping matrix $\bm{M}(k)$ of the Holstein-Primakoff bosons in the spin-wave Hamiltonian (\ref{h_lsw}) has the following form,
\begin{equation}
\bm{M}(k)=\left(
            \begin{array}{cc}
              \bm{A}(k) & \bm{B}(k) \\
              \bm{B}(k) & \bm{A}(k) \\
            \end{array}
          \right),
\end{equation}
where $\bm{A}(k)$ and $\bm{B}(k)$ are $L\times L$ matrices. The explicit forms of $\bm{A}(k)$ and $\bm{B}(k)$ depend on the magnetic order of the ground state.

(i) For the N\'{e}el order ($L=2$),
\begin{equation}
\bm{A}(k)=\left(
            \begin{array}{cc}
              a_{2}(k) & a_{1}^{\ast}(k) \\
              a_{1}(k) & a_{2}(k) \\
            \end{array}
          \right)
\end{equation}
and
\begin{equation}
\bm{B}(k)=\left(
            \begin{array}{cc}
              0 & a_{3}^{\ast}(k) \\
              a_{3}(k) & 0 \\
            \end{array}
          \right),
\end{equation}
in which
\begin{eqnarray}
a_1(k)&=\frac{1}{4}K(e^{i\bm{k}\cdot\mathbf{a}}-e^{i\bm{k}\cdot\bm{b}}),\quad a_2(k)=\frac{1}{2}(3J+K),\nonumber\\
a_3(k)&=(\frac{1}{2}J+\frac{1}{4}K)(e^{i\bm{k}\cdot\bm{a}}+e^{i\bm{k}\cdot\bm{b}})+\frac{1}{2}Je^{i\bm{k}\cdot\bm{c}}. \nonumber
\end{eqnarray}

(ii) For the zigzag order ($L=4$),
\begin{eqnarray}
\bm{A}(k)=\left(
            \begin{array}{cccc}
              a_{2}(k) & a_{1}(k) & 0 & 0 \\
              a_{1}^{\ast}(k) & a_{2}(k) & 0 & 0 \\
              0 & 0 & a_{2}(k) & a_{1}(k) \\
              0 & 0 & a_{1}^{\ast}(k) & a_{2}(k) \\
            \end{array}
          \right)
\end{eqnarray}
and
\begin{eqnarray}
\bm{B}(k)=\left(
            \begin{array}{cccc}
              0 & a_{3}(k) & 0 & a_{4}(k) \\
              a_{3}^{\ast}(k) & 0 & a_{4}^{\ast}(k) & 0 \\
              0 & a_{4}(k) & 0 & a_{3}(k) \\
              a_{4}^{\ast}(k) & 0 & a_{3}^{\ast}(k) & 0 \\
            \end{array}
          \right)
\end{eqnarray}
in which
\begin{eqnarray}
a_1(k)&=(\frac{1}{2}J+\frac{1}{4}K)(e^{i\bm{k}\cdot\bm{a}}+e^{i\bm{k}\cdot\bm{b}}),\quad
a_2(k)=\frac{1}{2}(K-J),\nonumber\\
a_3(k)&=\frac{1}{4}K(e^{i\bm{k}\cdot\bm{a}}-e^{i\bm{k}\cdot\bm{b}}),\quad
a_4(k)=\frac{1}{2}Je^{i\bm{k}\cdot\bm{c}}. \nonumber
\end{eqnarray}

(iii) For the stripy order ($L=4$),
\begin{eqnarray}
\bm{A}(k)=\left(
            \begin{array}{cccc}
              a_{2}(k)& a_{1}(k) & 0 & a_{4}(k) \\
              a_{1}^{\ast}(k) & a_{2}(k) & a_{4}^{\ast}(k) & 0 \\
              0 & a_{4}(k) & a_{2}(k) & a_{1}(k) \\
              a_{4}^{\ast}(k) & 0 & a_{1}^{\ast}(k) & a_{2}(k) \\
            \end{array}
          \right)
\end{eqnarray}
and
\begin{eqnarray}
\bm{B}(k)=\left(
            \begin{array}{cccc}
              0 & a_{3}(k) & 0 & 0 \\
              a_{3}^{\ast}(k) & 0 & 0 & 0 \\
              0 & 0 & 0 & a_{3}(k) \\
              0 & 0 & a_{3}^{\ast}(k) & 0 \\
            \end{array}
          \right)
\end{eqnarray}
in which
\begin{eqnarray}
a_{1}(k)&=\frac{1}{4}K(e^{i\bm{k}\cdot\bm{a}}-e^{i\bm{k}\cdot\bm{b}}),\quad
a_{2}(k)=\frac{1}{2}(J-K),\nonumber \\
a_{3}(k)&=(\frac{1}{2}J+\frac{1}{4}K)(e^{i\bm{k}\cdot\bm{a}}+e^{i\bm{k}\cdot\bm{b}}),\quad
a_{4}(k)=\frac{1}{2}J e^{i\bm{k}\cdot\bm{c}}. \nonumber
\end{eqnarray}

\section{Spin waves and their dependence on the interaction parameters}\label{spin-wave}
\setcounter{figure}{0}

\begin{figure}
  \centering
  \includegraphics[scale=1.1]{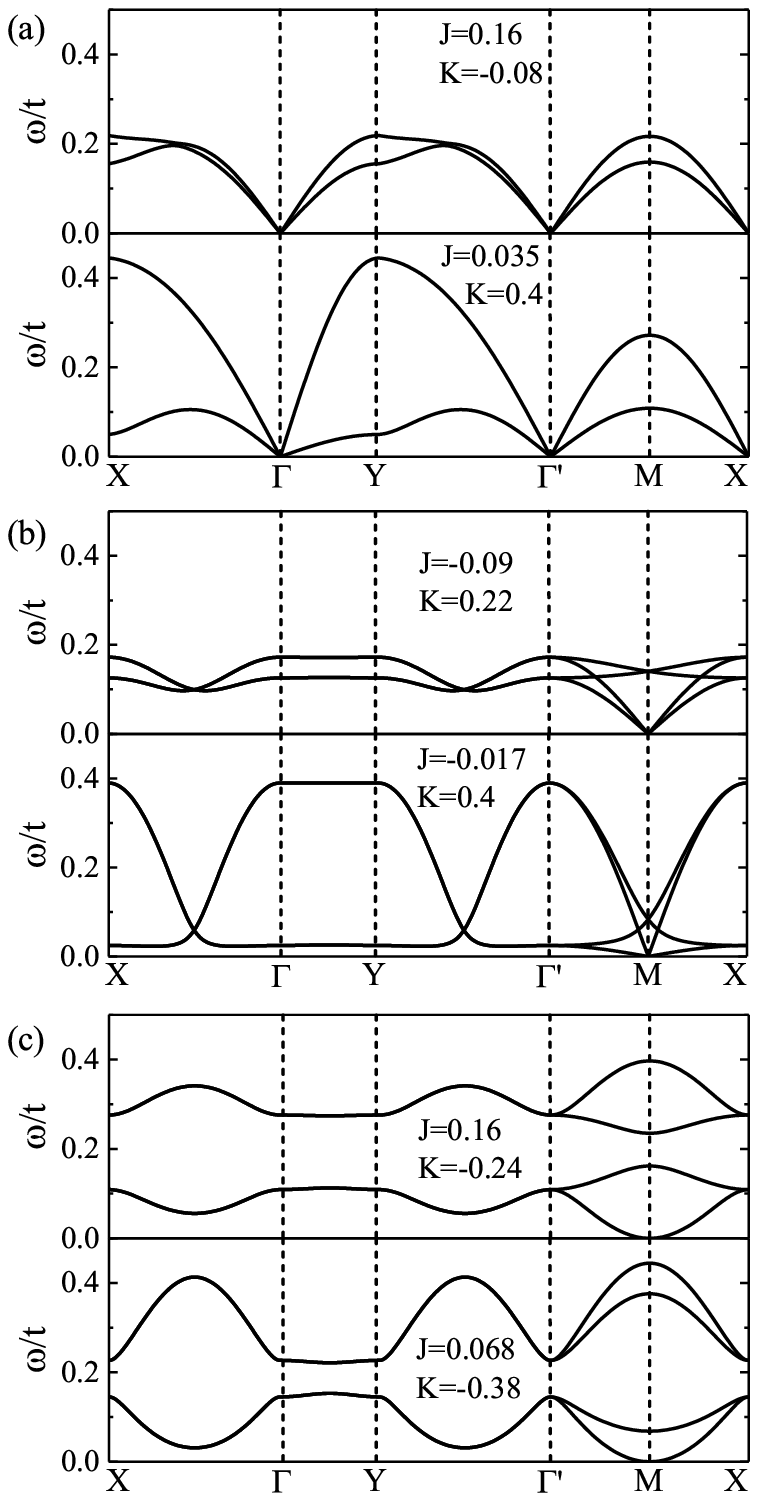}
  \caption{\label{fig:spin-wave}Spin-wave dispersions for (a) N\'{e}el AF, (b) zigzag and (c) stripy phases. In every phase, the results for two sets of parameters are presented.}
\end{figure}

Figure~\ref{fig:spin-wave} exhibits the spin waves in the N\'{e}el AF, zigzag and stripy phases. In each phase, we present the results for two sets of parameters, which are used in the main text. We can see that there is a common feature for the spin waves in all of the three magnetically ordered phases, i.e. the excitation energy of the lowest-energy branch in every phase is reduced with the increase of the Kitaev interaction $K$.

\section{Dispersions of bare holons and holon-magnon interaction vertices}\label{hole-magnon-vertex}

The operator $b_{k,\alpha}$ of the Holstein-Primakoff boson in equation~(\ref{h_lsw}) is related to the operator $\gamma_{k,n}$ of the magnon in equation~(\ref{eq:hdiagn}) through a generalized Bogoliubov transformation in the form of
\begin{equation}
b_{k,\alpha}=\sum_n\left[u_{\alpha n}(k)\gamma_{k,n} +v_{\alpha n}^{\ast}(-k)\gamma_{-k,n}^\dag\right],
\end{equation}
which can be obtained by diagonalizing the Hamiltonian (\ref{h_lsw}). Using this relation between $b_{k,\alpha}$ and $\gamma_{k,n}$, we can write the explicit expression of the bare holon dispersion $\bm{T}(k)$ and holon-magnon interaction $\bm{D}(k,q,n)$ in the Hamiltonian (\ref{hole-eq}).

(i) For the N\'{e}el order, $\bm{T}(k)$ is zero and
\begin{equation}
\bm{D}(k,q,n)=\left(
                 \begin{array}{cc}
                   0 & D_{12} \\
                   D_{21} & 0 \\
                 \end{array}
               \right),
\end{equation}
where
\begin{equation}
 D_{12}=t\sqrt{\frac{2}{N}}\left[u_{1n}^{\ast}(q)\sum_{\delta}e^{i\mathbf{(k-q)}\cdot\bm{\delta}}
 +v_{2n}^{\ast}(q)\sum_{\delta}e^{i\mathbf{k}\cdot\bm{\delta}}\right] \nonumber
\end{equation}
and
\begin{equation}
D_{21}=t\sqrt{\frac{2}{N}}\left[u_{2n}^{\ast}(q)\sum_{\delta}e^{-i(\bm{k}-\bm{q})\cdot\bm{\delta}}
 +v_{1n}^{\ast}(q)\sum_{\delta}e^{-i\bm{k}\cdot\bm{\delta}}\right] \nonumber
\end{equation}
with $\bm{\delta}=\bm{a},\bm{b},\bm{c}$.

(ii) For the zigzag order, we have
\begin{equation}
\bm{T}(k)=\left(
            \begin{array}{cccc}
              0 & T_{12} & 0 & 0 \\
              T_{21} & 0 & 0 & 0 \\
              0 & 0 & 0 & T_{34} \\
              0 & 0 & T_{43} & 0 \\
            \end{array}
          \right),
\end{equation}
in which $T_{12}=T_{34}=t(e^{-i\mathbf{k}\cdot\mathbf{a}}+e^{-i\mathbf{k}\cdot\mathbf{b}})$, $T_{21}=T_{12}^{\ast}$ and $T_{43}=T_{34}^{\ast}$. The hole-magnon interaction is
\begin{equation}
\bm{D}(k,q,n)=\left(
                \begin{array}{cccc}
                  0 & 0 & 0 & D_{14} \\
                  0 & 0 & D_{23} & 0 \\
                  0 & D_{32} & 0 & 0 \\
                  D_{41} & 0 & 0 & 0 \\
                \end{array}
              \right)
\end{equation}
where the four nonzero elements are
\begin{equation}
D_{23}=t\sqrt{\frac{4}{N}}\left[v_{3n}^{\ast}(q)e^{i\bm{k}\cdot\bm{c}}+u_{2n}^{\ast}(q)e^{i(\bm{k}-\bm{q})\cdot\mathbf{c}} \right],\nonumber
\end{equation}
\begin{equation}
D_{32}=t\sqrt{\frac{4}{N}}\left[v_{2n}^{\ast}(q)e^{-i\bm{k}\cdot\bm{c}}+u_{3n}^{\ast}(q)e^{-i(\bm{k}-\bm{q})\cdot\bm{c}} \right],\nonumber
\end{equation}
\begin{equation}
D_{14}=t\sqrt{\frac{4}{N}}\left[v_{4n}^{\ast}(q)e^{-i\bm{k}\cdot\bm{c}}+u_{1n}^{\ast}(q)e^{-i(\bm{k}-\bm{q})\cdot\bm{c}} \right],\nonumber
\end{equation}
and
\begin{equation}
D_{41}=t\sqrt{\frac{4}{N}}\left[v_{1n}^{\ast}(q)e^{i\bm{k}\cdot\bm{c}}
+u_{4n}^{\ast}(q)e^{i(\bm{k}-\bm{q})\cdot\bm{c}}\right]. \nonumber
\end{equation}

(iii) For the stripy order, we have
\begin{equation}
\bm{T}(k)=\left(
            \begin{array}{cccc}
              0 & 0 & 0 & T_{14} \\
              0 & 0 & T_{23} & 0 \\
              0 & T_{32} & 0 & 0 \\
              T_{41} & 0 & 0 & 0 \\
            \end{array}
          \right),
\end{equation}
in which $T_{23}=T_{41}=T_{32}^{\ast}=T_{14}^{\ast}=t e^{i\bm{k}\cdot\bm{c}}$. The hole-magnon interaction is
\begin{equation}
\bm{D}(k,q,n)=\left(
                \begin{array}{cccc}
                  0 & D_{12} & 0 & 0 \\
                  D_{21} & 0 & 0 & 0 \\
                  0 & 0 & 0 & D_{34} \\
                  0 & 0 & D_{43} & 0 \\
                \end{array}
              \right)
\end{equation}
where the nonzero elements are
\begin{equation}
D_{12}=t\sqrt{\frac{4}{N}}\left[u_{1n}^{\ast}(q)\sum_{\delta}e^{-i(\bm{k}-\bm{q})\cdot\bm{\delta}}+
 v_{2n}^{\ast}(q)\sum_{\delta}e^{-i\bm{k}\cdot\bm{\delta}}\right],
\nonumber
\end{equation}
\begin{equation}
D_{21}=t\sqrt{\frac{4}{N}}\left[u_{2n}^{\ast}(q)\sum_{\delta}e^{i(\bm{k}-\bm{q})\cdot\bm{\delta}}+
 v_{1n}^{\ast}(q)\sum_{\delta}e^{i\bm{k}\cdot\bm{\delta}}\right],
\nonumber
\end{equation}
\begin{equation}
D_{34}=t\sqrt{\frac{4}{N}}\left[u_{3n}^{\ast}(q)\sum_{\delta}e^{-i(\bm{k}-\bm{q})\cdot\bm{\delta}}+
 v_{4n}^{\ast}(q)\sum_{\delta}e^{-i\bm{k}\cdot\bm{\delta}}\right],
\nonumber
\end{equation}
and
\begin{equation}
D_{43}=t\sqrt{\frac{4}{N}}\left[u_{4n}^{\ast}(q)\sum_{\delta}e^{i(\bm{k}-\bm{q})\cdot\bm{\delta}}+
 v_{3n}^{\ast}(q)\sum_{\delta}e^{i\bm{k}\cdot\bm{\delta}}\right]
\nonumber
\end{equation}
with $\bm{\delta}=\bm{a},\bm{b}$.

\bibliography{single-hole}

\end{document}